\documentclass[conference]{IEEEtran}
\usepackage{graphicx,url}
\usepackage[caption=false,font=footnotesize, hangindent=10pt]{subfig}
\usepackage{amsmath,amsfonts,bm}
\usepackage[linesnumbered,lined,boxed,ruled]{algorithm2e}
\usepackage[hidelinks]{hyperref}

\graphicspath{{figure/}}
\newcommand{\E}{\mathbb{E}}
\newcommand{\argmax}{\mathop{\arg\max}}
\newcommand{\CI}{\mathcal{I}}
\newcommand{\CJ}{\mathcal{J}}

\begin{document}

\title{On Analyzing Estimation Errors due to Constrained Connections in Online Review Systems}

\author{
\IEEEauthorblockN{Junzhou Zhao}
\IEEEauthorblockN{Supplementary Information}
}

% use for special paper notices
%\IEEEspecialpapernotice{(Invited Paper)}

% make the title area
\maketitle

\begin{abstract}
In this work, we study how constrained connections can cause estimation 
errors in online review systems. Constrained connection is the phenomenon that 
a reviewer can only review a subset of products/services due to reviewer's narrow 
range of interests or limited attention capacity. We find that reviewers' 
constrained connections will cause poor inference performance, both from the 
measurements of estimation accuracy and Bayesian Cram\'er Rao lower bound. 

\end{abstract}

%\IEEEpeerreviewmaketitle

\section{Introduction} \label{sec:introduction}

Online reviews are more and more important factors for customers to decide whether 
to buy a product or service in online markets. Due to this reason, online review 
systems have become battle fields for companies to compete with each other by 
hiring ``Internet Water Mercenaries'', which are also known as paid spammers, to 
post favorable reviews about their products/services and negative reviews about 
their competitors'. These fake reviews disturb customers' judgments on the quality 
of products/services and ruin companies' reputation. Hence, an always important 
problem in online review systems is how to accurately obtain the \emph{truth} of 
both reviewers (e.g., the reviewer is a spammer or non-spammer) and items (e.g., 
the product/service is good or bad) according to unreliable online reviews. 

\begin{figure}
\includegraphics[width=\linewidth]{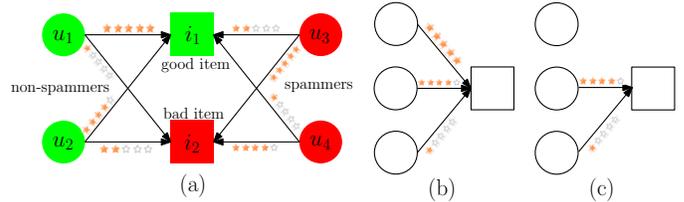}
\caption{Examples. \label{fig:example}}
\end{figure}

In previous studies\cite{Wang2011d,Akoglu2013}, most of the works ignore the 
function of the underlying topology of ORS. The topology of an 
online review system is a bipartite graph representing which reviewers can 
review which items. Many works explicitly or implicitly assume that reviewers can 
review all the other items, such as the example shown in Fig.~\ref{fig:example}(a). 
In fact, a reviewer can only review a subset of items in real-world, which results 
in \emph{constrained connections} for each reviewer in the topology. The 
constrained connections may be because of either the reviewer's narrow range of 
interests or the reviewer's limited attention capacity (that he cannot afford to 
review all other items). The constrained connections can affect the performance of 
jointly estimating the truth of reviewers and items. For example, let us consider a 
simplest online review system that consists of three reviewers and one item. If we 
assume the majority of reviewers are non-spammers (that is true in real-world), 
then in the case of Fig.~\ref{fig:example}(b), from this topology and reviews by 
each reviewer we can infer with high confidence that the item is probably good and 
the bottom reviewer is likely to be a spammer. However, in the case of (c), we 
cannot obtain a high confidence conclusion because we do not know the reviews of 
the top reviewer. 

The simple example tells us that different topologies of ORS 
along with unreliable reviews contain different amounts of information for jointly 
estimating the truth of reviewers and items. Actually, connections between 
reviewers and items act as constraints in such systems. They constrain the 
joint probability distribution of the truth of reviewer-item pairs they connect. 
For example, a non-spammer usually gives good (bad) items good (bad) reviews with 
high probability, which indicates that the truth of a reviewer and the truth of an 
item he reviewed are related. Hence the topology of the ORS yields a 
set of constraints that the truth of reviewers and items must obey, and these 
constraints help to reduce the uncertainty of parameters in the system. 

In order to compare the amounts of information contained in different topologies 
(and reviews), we calculate the Bayesian Cram\'er Rao lower bound (BCRLB) of 
maximum a 
posteriori estimator (MAPE) in such systems for different bipartite graph models. 
We find that BCRLB varies for different topologies. This indicates that for some 
topologies the truth become much difficult to be estimated by any MAPEs.

\section{Background and Basic Results} \label{sec:estimation}

\subsection{Data Model}

Following the existing works\cite{Raykar2010,Karger2011,Liu2012}, we assume that there are a set of reviewers $V$ and a set of items $I$ in an online review system. Each item $i\in I$ is associated with a binary label $z_i\in\{\pm 1\}$, which is considered to be a random variable representing the quality of item $i$, e.g., $z_i=+1$ if item $i$ is \emph{good}; $z_i=-1$ if $i$ is \emph{bad}. Each reviewer can choose items to review. A review represents the reviewer's attitude to an item. If we use $r_{ui}\in \{\pm 1\}$ to denote $u$'s review to $i$, then $r_{ui}=+1$ (or $r_{ui}=-1$) means that reviewer $u$ considers item $i$ to be good (or bad). However, reviewers are not always accurate to review items, and we use $\theta_u\in [0,1]$ to represent the probability that the reviewer can give correct reviews, i.e., $\theta_u=P(r_{ui}=z_i)$. In practice, it is reasonable to assume that the majority of reviewers have $\theta_u>0.5$. This is achieved by putting a prior distribution on $\theta_u$. A nature choice of such a prior is the beta distribution, i.e., $P(\theta_u)\propto \theta_u^{\alpha-1} (1-\theta_u)^{\beta-1}$, where $\alpha>\beta$. 

Be different from previous works, in this work we assume that a reviewer can not 
freely choose which items to review. The reasons may be the reviewer's narrow range 
of interests, limited attention capacity, and so on. If reviewer $u$ can review 
item $i$, we connect $u$ and $i$ by an edge $(u,i)$. This forms a bipartite graph 
$G(V,I,E)$, where $E$ is the set of edges. Furthermore, we use $I_u$ to denote the 
set of items that $u$ can review, and we use $V_i$ to denote the set of reviewers 
who can review $i$. 

To make the aforementioned model more general, we assume that items to be reviewed 
by reviewers are chosen independently \emph{with replacement} constrained by graph 
$G$\footnote{Consider items to be shops, each time a consumer buy a product from a 
shop, he can review the shop.}. This forms a collection of $n$ review samples 
$R=\{r_1,r_2,\cdots,r_n\}$ where $r_k$ denotes the $k$-th sample representing some 
reviewer $u$ gives some item $i$ a review $r_{ui}$. Since items are chosen with 
replacement, we may observe that reviewer $u$ reviews item $i$ many times. We use 
$n_{ui}^x$ to represent the number of times $u$ gives $i$ a review $x$ in the 
samples $R$. Note that $n_{ui}^x$ satisfies $\sum_{u\in V}\sum_{i\in 
I}\sum_{x\in\{\pm 1\}}n_{ui}^x=n$. 

Our goal is to study how $G$ can affect the estimation accuracy when using $R$ to estimate $\theta=\{\theta_u\}_{u\in V}$ and $z=\{z_i\}_{i\in I}$.

\subsection{Maximum A Posteriori Estimator}

A convenient way to estimate parameters of the previous model is by considering $\theta$ as parameters and $z$ as hidden variables\cite{Dawid1979}. David and Skene\cite{Dawid1979} presented an expectation maximization (EM) approach to maximize the likelihood. Here we propose to maximize the posteriori of $\theta$ which can include the priori information of $\theta$. That is, 
\begin{equation}
\max\log P(\theta|R)=\max\log\sum_z P(\theta,z|R).
\end{equation}

\noindent\textbf{E-Step:} In the E-Step, we need to calculate the probability of hidden variables given the other variables $P(z|R,\theta)$, which can be factorized to $\prod_iP(z_i|R_{\cdot i},\theta)$. Here $R_{\cdot i}\subseteq R$ denotes the reviews in the samples that are related to item $i$. If we denote each factor by $\mu_i(z_i)$, then we can obtain
\begin{align}
\mu_i(z_i) \equiv&  P(z_i|R_{\cdot i},\theta) 
= \frac{P(R_{\cdot i}|z_i,\theta)P(z_i|\theta)}{P(R_{\cdot i}|\theta)} \\
\propto& P(z_i)\prod_{u\in V_i}P(R_{ui}|z_i,\theta_u) \\
=& P(z_i)\prod_{u\in V_i}P(r_{ui}=z_i|z_i,\theta_u)^{n_{ui}^{z_i}} \\ 
&\times P(r_{ui}=-z_i|z_i,\theta_u)^{n_{ui}^{-z_i}} \\
=& P(z_i)\prod_{u\in V_i}\theta_u^{n_{ui}^{z_i}} (1-\theta_{ui})^{n_{ui}^{-z_i}}. 
\end{align}

\noindent\textbf{M-Step:} In the M-Step, we need to solve 
\begin{align}
\theta^{(t+1)} 
&= \argmax_\theta Q(\theta,\theta^{(t)}) \\
&= \argmax_\theta\E_{z|R,\theta^{(t)}}\left[\log P(\theta,z|R)\right] \\
&= \argmax_\theta\E_{z|R,\theta^{(t)}}\left[\log P(R|\theta,z)+\log P(\theta) \right],
\end{align}
which gives us the following result
\begin{equation}
\theta_u^{(t+1)}=\frac{\sum_{i\in I_{u}}\sum_{x\in\{\pm 1\}}n_{ui}^{x}\mu_{i}(x)+\alpha-1}{|R_{u\cdot}|+\alpha+\beta-2}.
\end{equation}
Here, $R_{u\cdot}$ is the set of reviews given by reviewer $u$. 

The E-step and M-step of the EM algorithm implicitly defines an estimator of 
$\theta$, i.e., $\hat{\theta}_\text{MAP}=\text{EM}(R)$. Since $R$ is related to 
$G$, then $\hat{\theta}_\text{MAP}$ is also related to $G$. To understand how $G$ 
can affect the MAP estimator, we go to study the Mean Squared Errors of 
$\hat{\theta}_\text{MAP}=\{\hat{\theta}_u\}_{u\in V}$.

\section{Estimation Errors Analysis} \label{sec:fisherinformation}

\subsection{Lower Bound on Estimation Errors}

The Mean Squared Error of $\hat{\theta}_u$ is defined as $\text{MSE}(\hat{\theta}_u)=\E[\hat{\theta}_u-\theta_u]^2$, which is lower bounded by the Bayesian Cram\'er Rao lower bound (BCRLB) under some conditions\cite[Chapter 2]{VanTrees1968}. We rewrite Eq.~(217) in Van Trees' book \cite[Page 73]{VanTrees1968} and obtain the following relationship
\begin{equation}
\text{MSE}(\hat{\theta}_u) \geq [\CJ^{-1}]_{uu},
\end{equation}
where 
\begin{equation}
\CJ_{uv} = -\E\left[\frac{\partial^{2}\log P(\theta|R)}{\partial\theta_u\partial\theta_v}\right]
\end{equation}
is the element $(u,v)$ of Fisher information matrix $\CJ$. 

The above relationship requires that $\hat{\theta}_\text{MAP}$ is \emph{weakly unbiased}\cite[Chapter 2]{VanTrees1968}, which is unknown for the MAP estimator defined by EM algorithm. However, it is known that under general conditions, for large $n$, the posterior distribution of $\theta$ can be approximated by normal distribution\cite{Cramer1946,VanTrees1968,Efron1978,Tanner2006}
\[
P(\theta|R)\rightarrow\mathcal{N}(\hat{\theta}_\text{MAP},\CI(\hat{\theta}_\text{MAP})^{-1})\text{ as } n\rightarrow\infty,
\]
where $\CI({\hat{\theta}}_\text{MAP})$ is the {\em observed Fisher information matrix}, and each element $(u,v)$ of $\CI({\hat{\theta}}_\text{MAP})$ is defined by
\[
[\CI(\hat{\theta}_\text{MAP})]_{uv}=-\frac{\partial^{2}\log P(\theta|R)}{\partial\theta_u\partial\theta_v}\biggr|_{\theta=\hat{\theta}_\text{MAP}}.
\]

The above conclusion tells us that $\hat{\theta}_\text{MAP}$ defined by the EM algorithm is a consistent estimator of $\theta$ with covariance matrix determined by $\CI$. For different $G$'s, the estimator $\hat{\theta}_\text{MAP}$ will have different covariance matrices. We can compare the estimation errors by evaluating $\CI$'s on different bipartite graphs. In the following, we find that $\CI$ is a diagonal matrix and it can be efficiently computed in combining with the EM procedure. 

\subsection{Obtaining BCRLB in Combining with EM Procedure}

Because $P(\theta|R)P(z|\theta,R)=P(\theta,z|R)$, or equivalently
\begin{equation}
\log P(\theta|R)=\log P(\theta,z|R)-\log P(z|\theta,R), 
\end{equation}
Then 
\begin{align}
\frac{\partial^{2}\log P(\theta|R)}{\partial\theta_u\partial\theta_v}
=&\frac{\partial^{2}\log P(\theta,z|R)}{\partial\theta_u\partial\theta_v}-\frac{\partial^{2}\log P(z|\theta,R)}{\partial\theta_u\partial\theta_v} \\
=&\sum_z\frac{\partial^{2}\log P(\theta,z|R)}{\partial\theta_u\partial\theta_v}P(z|\theta^{(t)},R) \\ &-\sum_z\frac{\partial^{2}\log P(z|\theta,R)}{\partial\theta_u\partial\theta_v}P(z|\theta^{(t)},R) \\
\equiv&\frac{\partial^{2}Q(\theta,\theta^{(t)})}{\partial\theta_u\partial\theta_v}- \frac{\partial^{2}H(\theta,\theta^{(t)})}{\partial\theta_u\partial\theta_v} 
\end{align}
The first item of RHS is 
\begin{align}
\frac{\partial^2Q}{\partial\theta_u^2}
=&\sum_{i\in I_u}\sum_{x\in\{\pm 1\}}\mu_i(x)\left[-\frac{n_{ui}^x}{\theta_u^2} -\frac{n_{ui}^{-x}}{(1-\theta_u)^2}\right] \\
&-\frac{\alpha-1}{\theta_u^2}-\frac{\beta-1}{(1-\theta_u)^2}
\end{align}
and $\frac{\partial^2Q}{\partial\theta_u\partial\theta_v}=0$ if $u\neq v$.
The second item of RHS is
\begin{align}
\frac{\partial^2H}{\partial\theta_u^2}
=& \sum_{i\in I_u}\sum_{x\in\{\pm 1\}}\mu_i(x)\left[-\frac{n_{ui}^x}{\theta_u^2}-\frac{n_{ui}^{-x}}{(1-\theta_u)^2}\right], 
\end{align}
and $\frac{\partial^2H}{\partial\theta_u\partial\theta_v}=0$ if $u\neq v$.
Finally, we obtain the observed Fisher information matrix
\begin{equation}
\CI_{uu}=\frac{\alpha-1}{\hat{\theta}_u^2}+\frac{\beta-1}{(1-\hat{\theta}_u)^2},
\label{eq:Iuu}
\end{equation}
and $\CI_{uv}=0$ if $u\neq v$. 

This indicates that $\CI$ is a diagonal matrix. Note that Eq.~\eqref{eq:Iuu} is convex, $\CI_{uu}$ gets the minimum value at $\hat{\theta}_u^*=\frac{1}{1+\sqrt[4]{(\beta-1)/(\alpha-1)}}$ and $\CI_{uu}$ gets the maximum value at $0$ or $1$. This tells us that $\hat{\theta}_u$ is most \emph{uncertain} when $\hat{\theta}_u=\hat{\theta}_u^*$ and most \emph{certain} at $\hat{\theta}_u=0$ or $1$. This is consistent with intuition as $\hat{\theta}_u$ can be considered as the parameter of a Bernouli distribution.

\section{Empirical Results} \label{sec:experiment}

To study how constrained connections can affect the estimation accuracy of MAPE, we 
first present several bipartite graph models and then study how these models affect 
the performance of MAPE measured by the accuracy of classifying items and BCRLBs. 

\subsection{Bipartite Graph Models}

\subsubsection{Random Graph Model $G_\text{rnd}$}

Each edge $(u,i)$ in $G_\text{rnd}$ is formed by uniformly choosing a reviewer $u\in V$ and uniformly choosing an item $i\in I$.

\subsubsection{Item Preferential Attachment Graph Model $G_\text{iPA}$}

The assumption of this model is that popular items are more easily to receive reviews. Hence, an edge $(u,i)$ in $G_\text{iPA}$ is formed by uniformly random choosing a reviewer $u\in V$, and choosing item $i\in I$ with probability proportion to $i$'s degree in $G_\text{iPA}$.

\subsubsection{Reviewer and Item Preferential Attachment Graph Model $G_\text{riPA}$}

We can also assume that a reviewer who is more active is more likely to review items. Hence, an edge $(u,i)$ in $G_\text{iPA}$ is formed by choosing a reviewer $u\in V$ with probability proportion to $u$'s degree, and choosing item $i\in I$  with probability proportion to $i$'s degree in $G_\text{iPA}$.

\subsection{Building Ground Truth Known Datasets}

Given a graph built by one of the above models, we describe the procedure of generating review samples $R$. 

We specify a set of $|V|$ reviewers and $|I|$ items. Suppose that each user $u$'s parameter $\theta_u$ is chosen from beta prior distribution $P(\theta_u)\propto \theta_u^{\alpha-1}(1-\theta_u)^{\beta-1}$, i.e., reviewer $u$ gives correct review with prior probability $\alpha/(\alpha+\beta)$. For each item $i$, we randomly assign a label $z_i\in \{\pm1\}$ by flipping a fair coin, i.e., $P(z_i=+1)=P(z_i=-1)=0.5$. The procedure of generating $R$ is as follows. 

\begin{algorithm}
\caption{Generating $R$.}
\KwIn{$G(V,I,E),\{z_i\}_{i\in I},\{\theta_u\}_{u\in V},n.$}
\KwOut{Review samples $R$.}
$R=[]$\;
\While{$|R|<n$,}{
	Randomly choose an edge $(u,i)\in E$\;
	Generate a random number $x\sim U(0,1)$\;
	$r_{ui}=z_i$ if $x\leq \theta_u$ else $-z_i$\;
	Put $r_{ui}$ into $R$\;
}
\end{algorithm}

\subsection{Comparing Items Inference Accuracy Under Different Graphs}

In the first experiment, we compare classification accuracy of items under different graph models. We set an item with label $+1$ (or $-1$) if $\mu_i(+1)>0.5$ (or $\mu_i(-1)>0.5$). The accuracy is defined as
\[
\text{Accuracy}=\frac{TP+TN}{P+N},
\]
where $TP$ and $TN$ are the \emph{true positive} and \emph{true negative} respectively. $P$ and $N$ are \emph{positive} and \emph{negative} respectively. Accuracy describes the fraction of items that can be corrected inferred. 

\begin{figure*}[t]
\centering
\subfloat[$|E|=1000$]{\includegraphics[width=.33\linewidth]{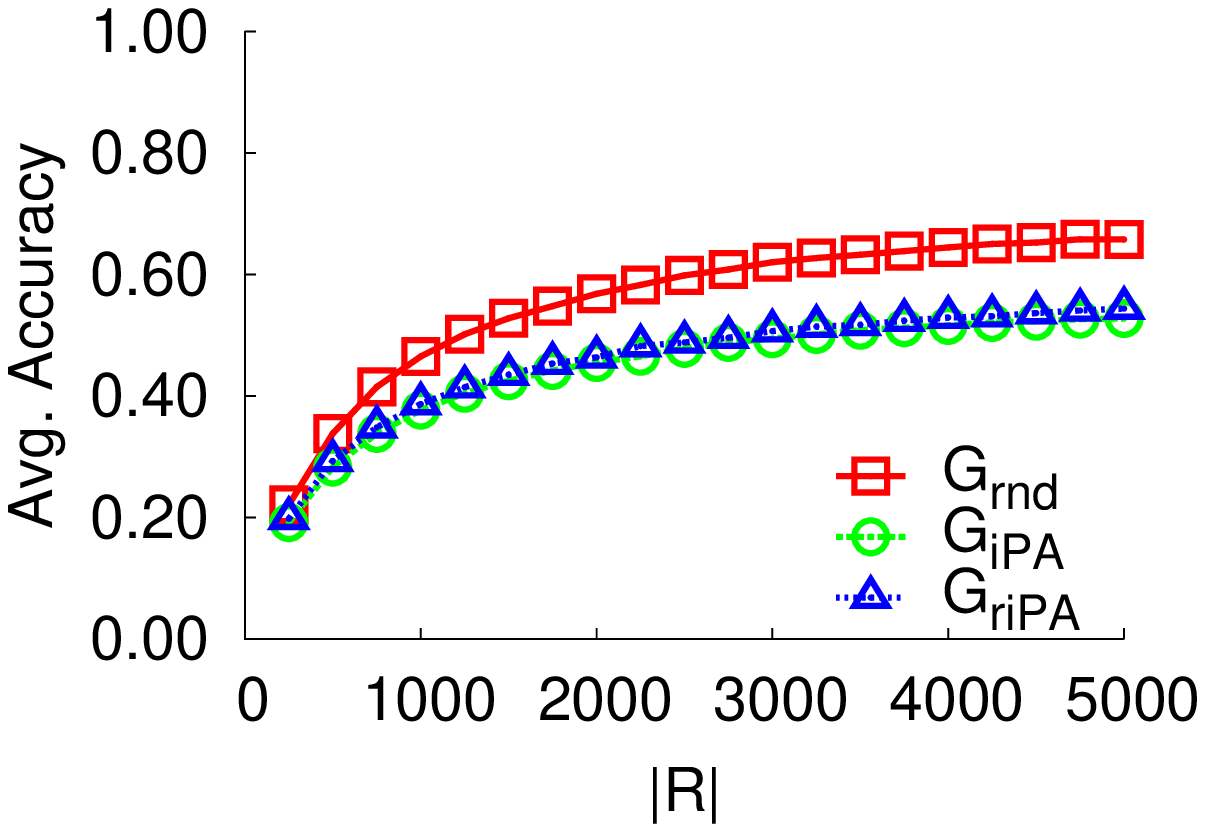}}
\subfloat[$|E|=2000$]{\includegraphics[width=.33\linewidth]{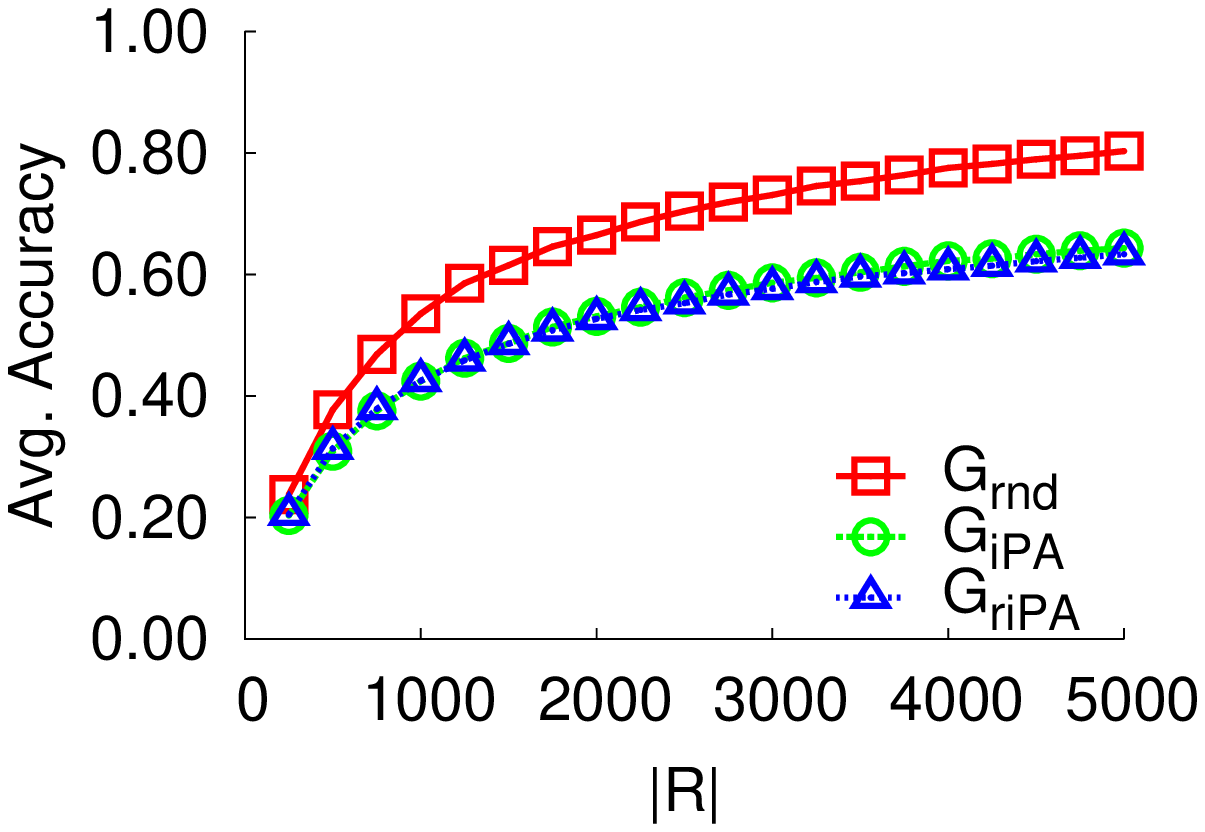}}
\subfloat[$|E|=3000$]{\includegraphics[width=.33\linewidth]{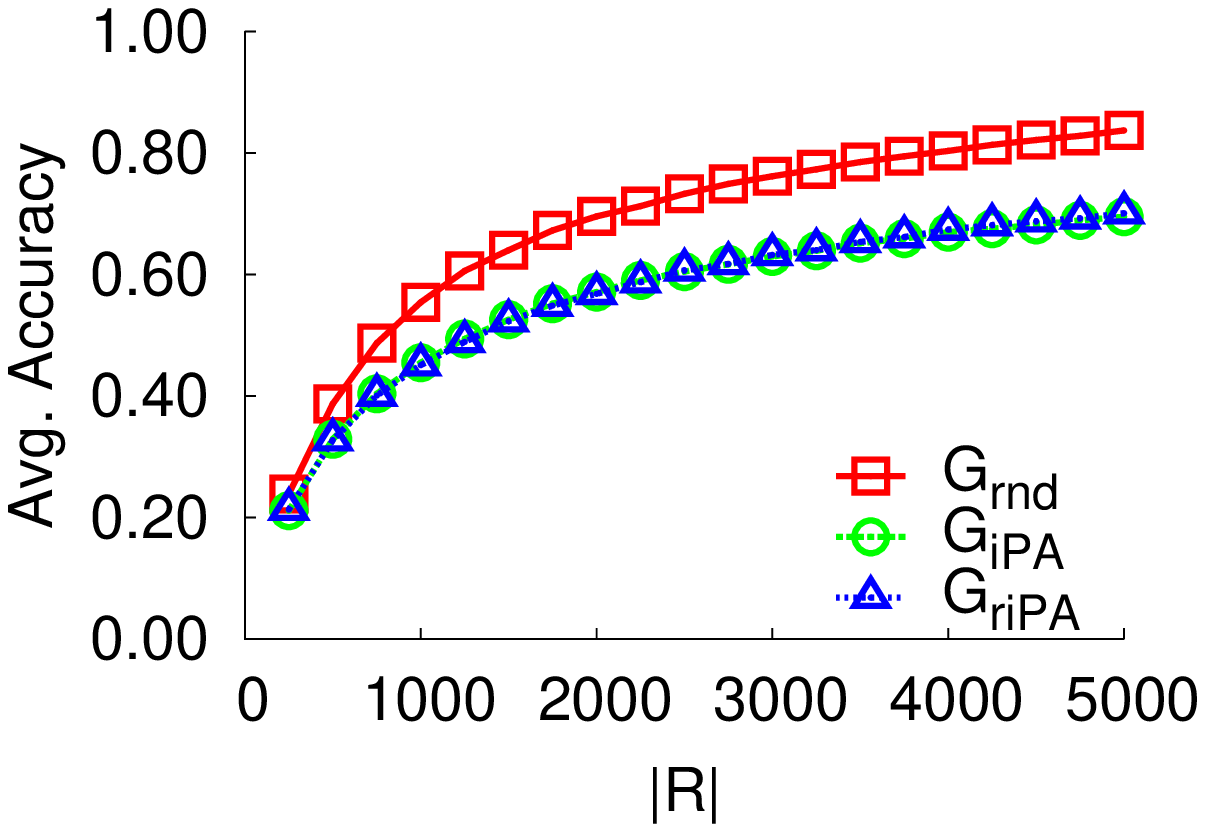}}\\
\subfloat[$|E|=4000$]{\includegraphics[width=.33\linewidth]{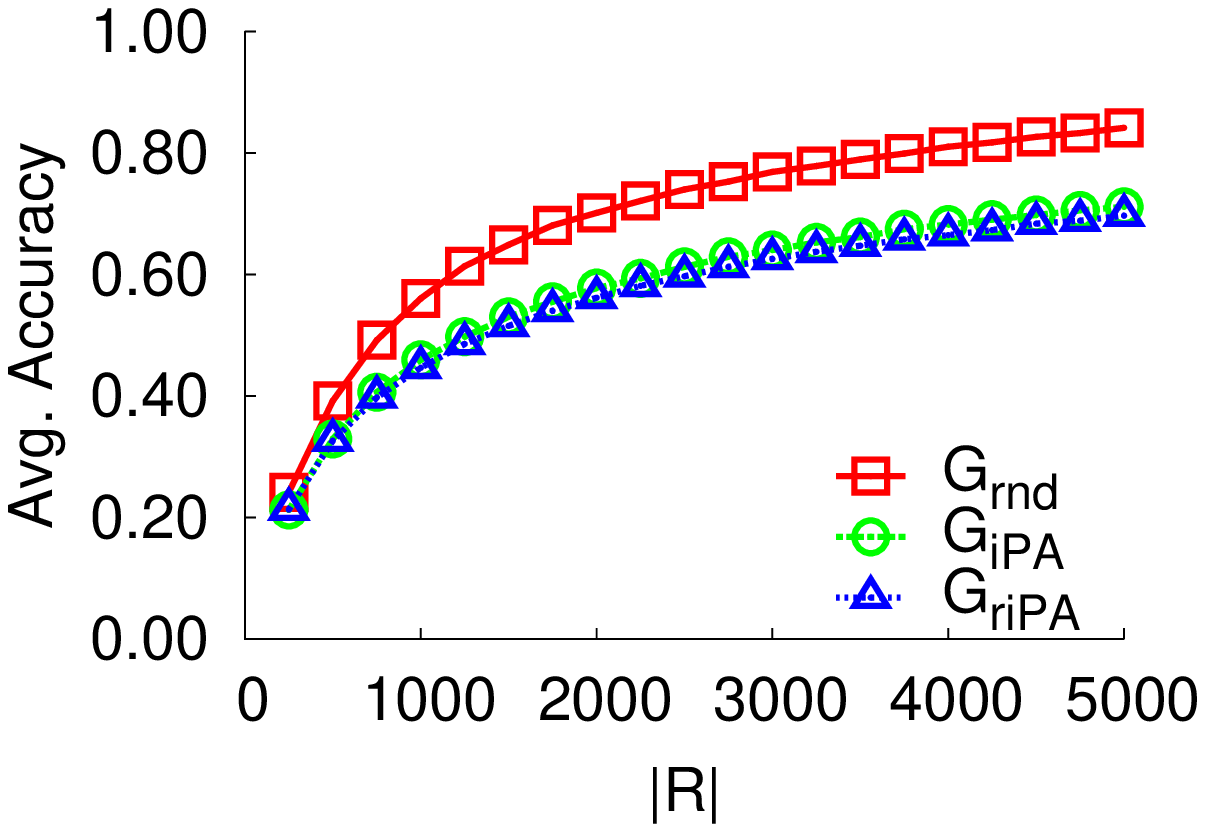}}
\subfloat[$|E|=5000$]{\includegraphics[width=.33\linewidth]{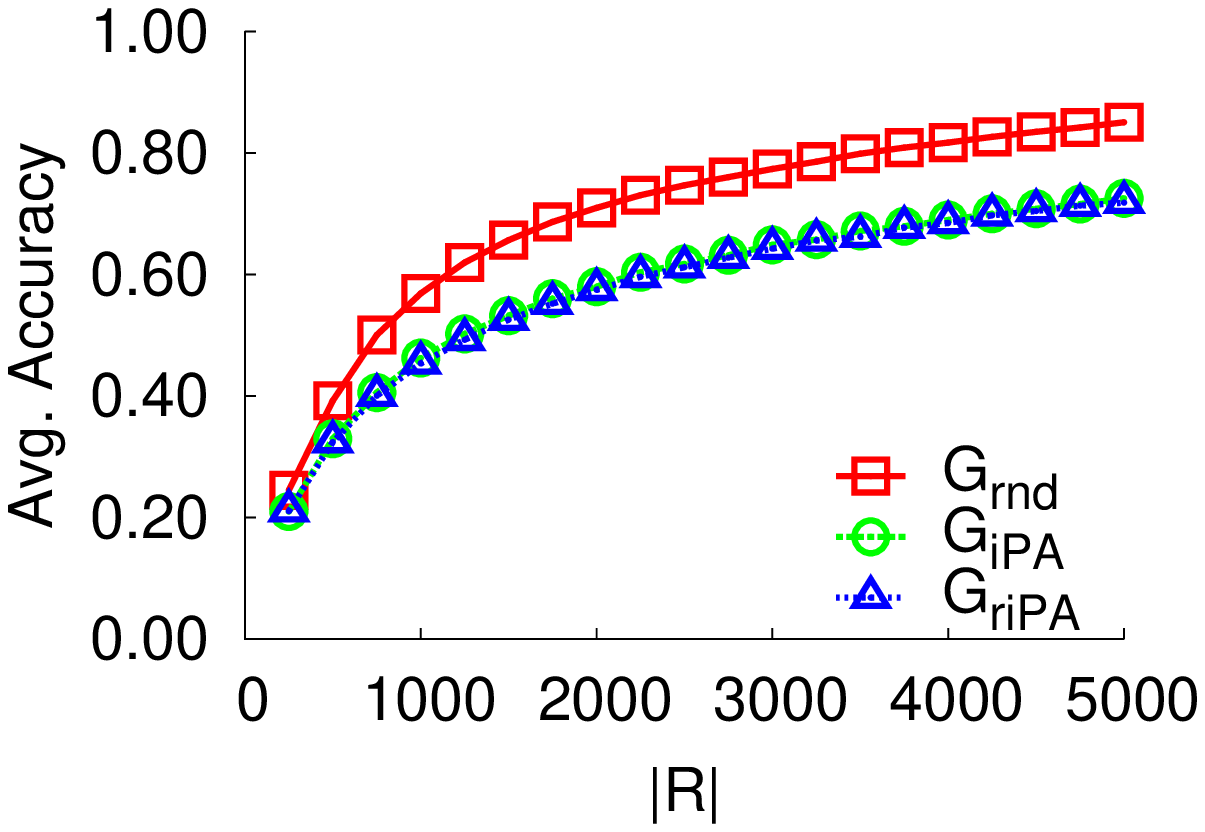}}
\caption{Items classification accuracy comparison. \label{fig:acc}}
\end{figure*}

The results are shown in Fig.~\ref{fig:acc}. We first generated graphs with number of nodes $|V|=500$ and varying number of edges ($|E|=1000,2000,3000,4000,5000$) using different graph models. In each figure, we generated review samples of different sizes ($500\leq |R|\leq 5000$), and show accuracy of inferring items averaged over 100 experiments respectively. We observe that when $|R|$ increases, the accuracy also increases and approaches 1. This confirms that the MAPE estimator is asymptotically unbiased. For different graph models, we observe that the accuracy on $G_\text{rnd}$ is larger than the other two models. This indicates that constrained connections will make the inference performance poor. However, the accuracy curves on $G_\text{iPA}$ and $G_\text{riPA}$ are approximately the same. This indicates that more constrained may not always decrease accuracy. To distinguish the difference of different constrained connections clearly, we study their difference of BCRLBs. 

\subsection{Comparing Estimation Errors Under Different Graphs}

\begin{figure*}[t]
\centering
\subfloat[$|E|=1000$]{\includegraphics[width=.33\linewidth]{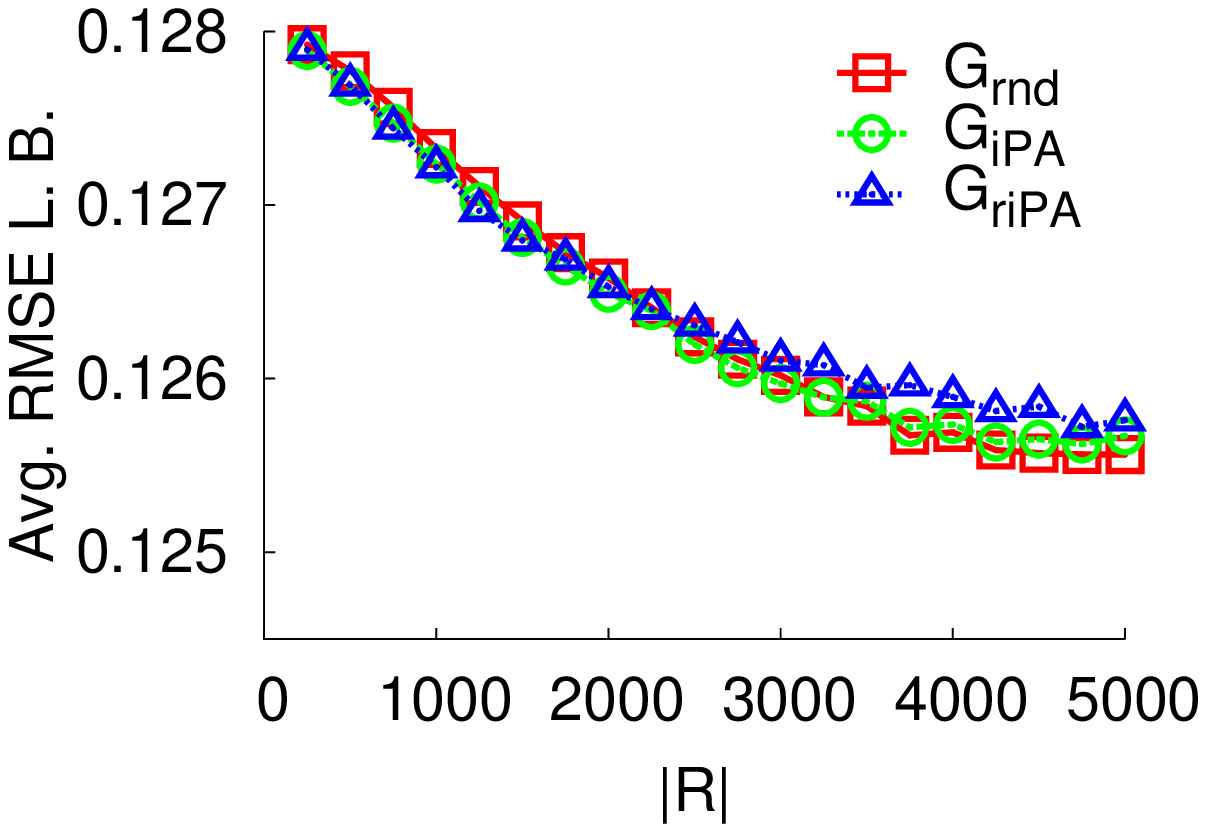}}
\subfloat[$|E|=2000$]{\includegraphics[width=.33\linewidth]{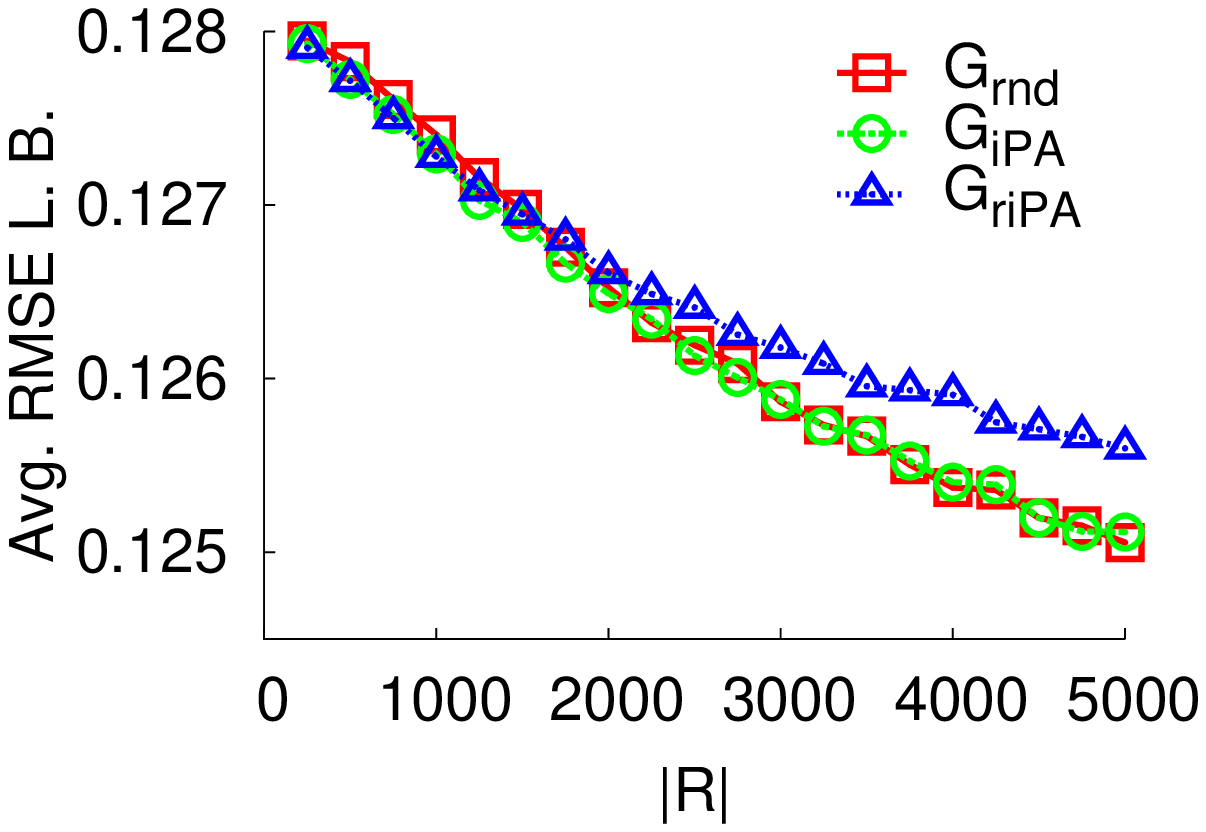}}
\subfloat[$|E|=3000$]{\includegraphics[width=.33\linewidth]{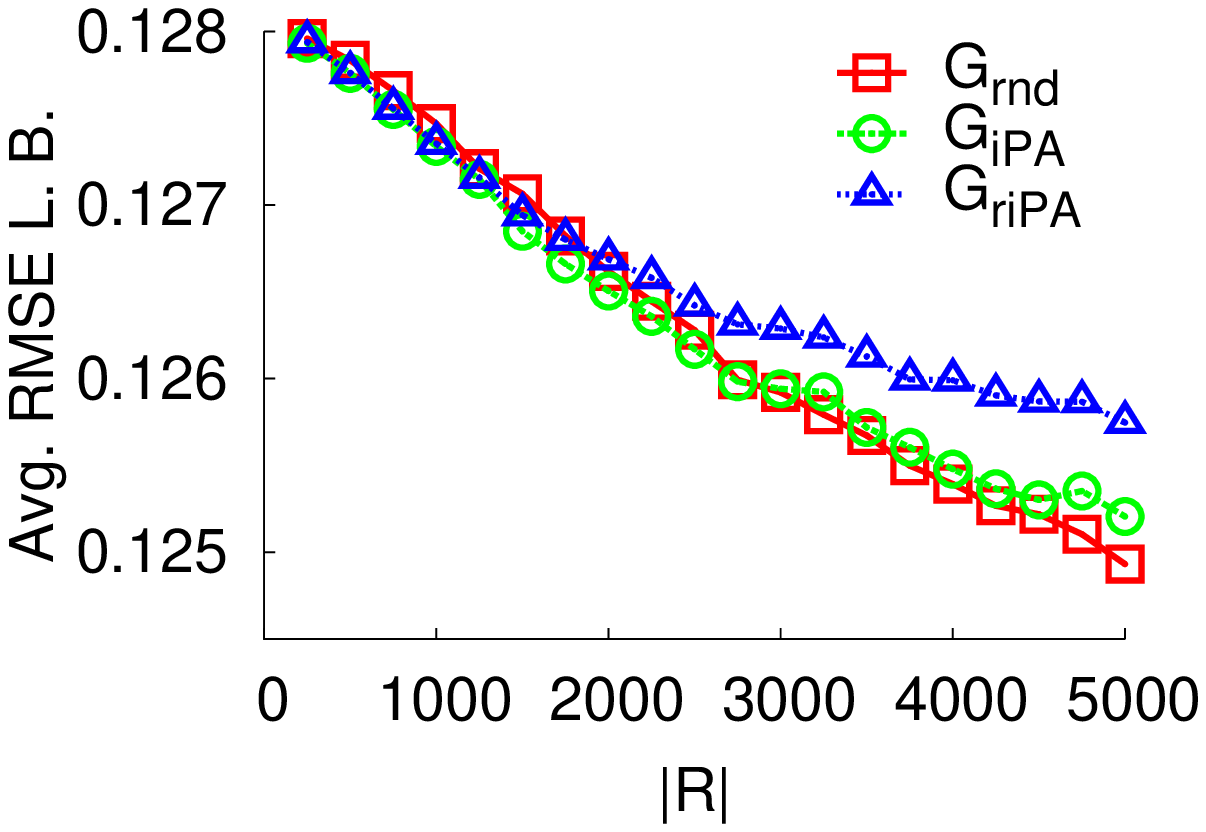}}\\
\subfloat[$|E|=4000$]{\includegraphics[width=.33\linewidth]{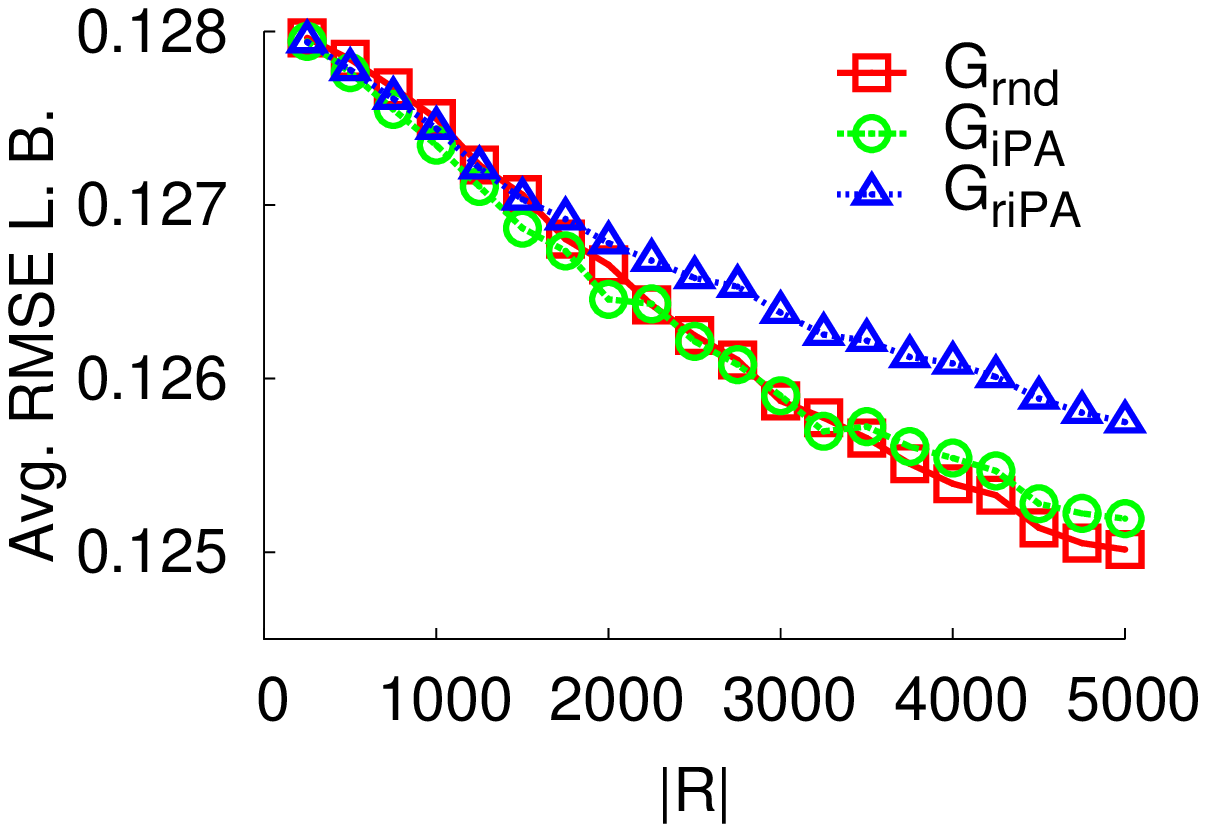}}
\subfloat[$|E|=5000$]{\includegraphics[width=.33\linewidth]{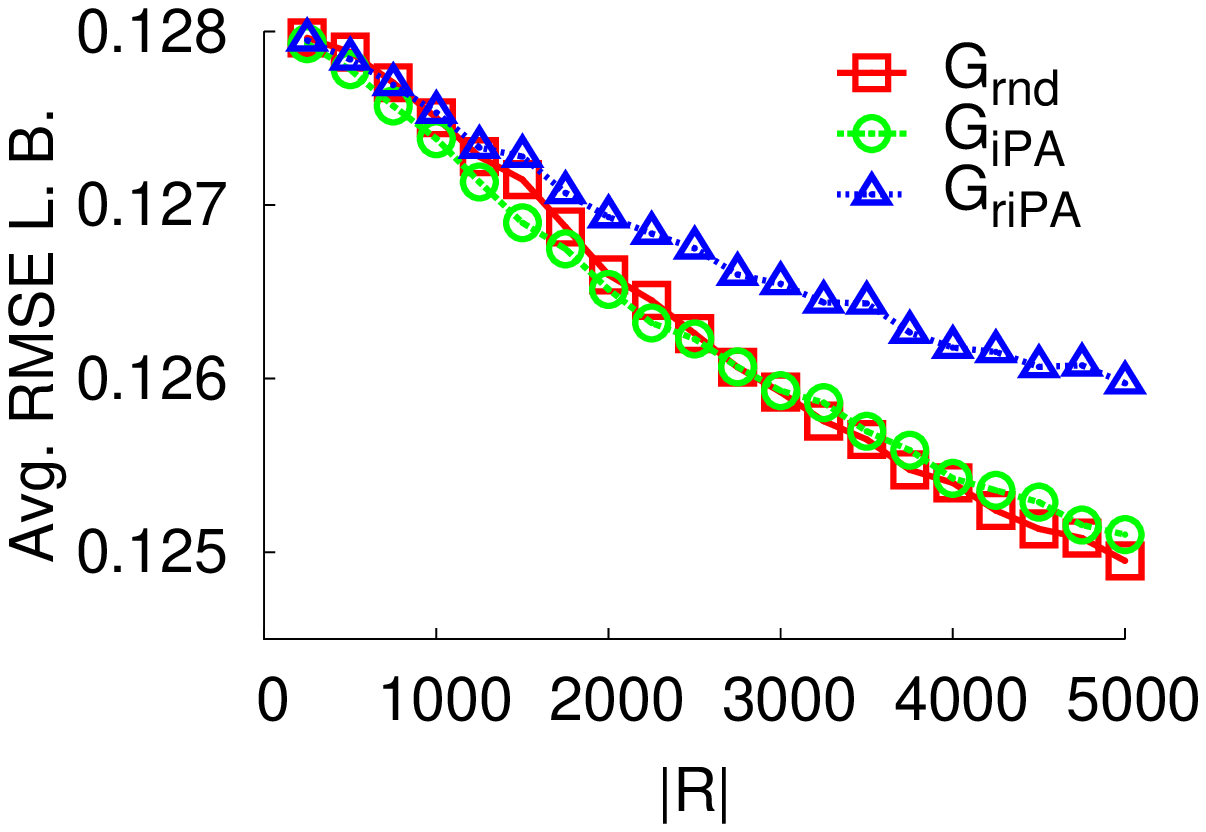}}
\caption{Estimation error comparison. \label{fig:rmse}}
\end{figure*}

In the second experiment, we study how different graph modes affect the BCRLBs. The settings are same as in the previous experiment. We compare the average rooted mean squared error (RMSE) (defined as $\text{RMSE}=\sqrt{\text{MSE}}$) lower bound over different graph models in Fig.~\ref{fig:rmse}. 

The RMSE decreases approximately with rate $1/n$ over all the graphs. For different graphs, when $n$ is large (we do not consider BCRLB for small n, because MAPE is biased when $n$ is small), RMSE on $G_\text{riPA}$ has the largest lower bound, then comes $G_\text{iPA}$ and RMSE on $G_\text{rnd}$ has the lowest lower bound. This indicates, when more constraints are added on graphs, the RMSE of any MAPEs will always become worse. 

\section{Conclusion} \label{sec:conclusion}

The constrained connections are common in real world. A reviewer cannot review all the items due to various reasons in online review systems. In this study, we find that this constrained connection will always cause poor inference performance, both from the viewpoints of inference accuracy and RMSE lower bound. 

\bibliographystyle{plain}
\bibliography{reference}

% that's all folks
\end{document}